\documentclass[12pt,preprint]{aastex}
\usepackage[dvips]{color}
\usepackage{graphicx}
\usepackage{amsmath}
\usepackage{amssymb}
\usepackage{natbib}
\bibpunct[,]{(}{)}{;}{a}{,}{,}   
\newcommand{\HI}{\mbox{\sc      H{i}}}
\newcommand{\HII}{\mbox{\sc      H{ii}}}

\slugcomment{The Evolving ISM in the Milky Way \& Nearby Galaxies}

\shorttitle{Star-forming regions in Virgo galaxies}
\shortauthors{Wong \& Kenney}

\begin{document}


\title{A Spitzer study of star-forming regions in Virgo Cluster galaxies}


\author{O.\ Ivy Wong\altaffilmark{1}, Jeffrey D. P. Kenney\altaffilmark{1}}
\affil{$^{1}$Astronomy Department, Yale University, P.O. Box 208101
    New Haven, CT 06520-8101}
\email{ivy.wong@yale.edu}


\begin{abstract}
We present a preliminary study of the star formation distribution 
within three Virgo Cluster galaxies  using the 24 $\mu$m Spitzer 
observations from the Spitzer Survey of Virgo (SPITSOV) in combination
with H$\alpha$ observations.  
The purpose of our study is to explore the relationship between the star formation 
distribution within galaxies and the type (and phase)
of interactions experienced within the cluster environment.  Neither highly-obscured 
star formation nor strongly enhanced star-forming regions along the leading edges of 
galaxies experiencing ICM--ISM interactions were found.  However, very unobscured 
star-forming regions were found in the outer parts of one galaxy (NGC 4402), while
 relatively obscured star-forming regions were found in the extraplanar regions
of another galaxy (NGC 4522). We attribute the observed differences between NGC 4402
and NGC 4522  to the direction of motion of each galaxy through the ICM.

\end{abstract}



\section{Introduction}
Currently, the general understanding of galaxy evolution suggests that
interactions play a significant role.  However, the dominant processes
are still not clear.  Galaxy clusters provide a good laboratory in which
to examine the effects of interactions on the evolution of individual
galaxies within the cluster.  The systematic differences in optical 
morphology between galaxies observed in the dense cores of clusters and 
those from the field regions \citep[i.e.\ the morphology--density 
relation ][]{dressler80} suggest that galaxy evolution is strongly driven 
by the environment.  Effects of the various forms of interactions
have been observed in detail in the Virgo Cluster where it is possible
to spatially resolve regions of dust, gas and stars in individual galaxies.


\citet{koopmann04} used the spatial distribution of H$\alpha$ to study 
the environmental effects on star formation in the Virgo Cluster galaxies.
However, H$\alpha$ observations are compromised by dust extinction
effects.  Without accounting for dust extinction effects, it is difficult to 
answer important questions about star formation such as:  i) the existence of 
enhanced star-forming regions due to ram pressure or gravitational effects,
and ii) the comparison between extraplanar and disk star formation rates.
Recently, it has been shown that a combination of H$\alpha$ and 24 $\mu$m
emission has been found to be a reliable indicator of total star formation
\citep{calzetti07}, since the 24 $\mu$m emission is presumed to consist 
mostly of re-radiated dust emission from the extincted star-forming regions.
Hence, we intend to use a combination of H$\alpha$ and 24 $\mu$m observations
to explore how the type and phase of interaction experienced relates to the 
observed spatial variations of star formation within galaxies and the global
variations between galaxies.

We present the  preliminary study of three Virgo 
galaxies known to be three of the best examples of cluster galaxies 
experiencing ram pressure; NGC 4522, NGC 4402 and NGC 4501.  
 Our preliminary study of these three galaxies  
is discussed in Section 2. Section 3 provides a summary.

\section{Preliminary study on NGC 4522, NGC 4402 and NGC 4501}
NGC 4522, NGC 4402 and NGC 4501 are three of the best examples of ram 
pressure stripping in action \citep{kenney04,crowl05,vollmer08,vollmer07}. 
 The \HI\ deficiencies of these 
three galaxies indicate that these galaxies have approximately three to four times
less \HI\ as one would expect from a `normal' galaxy.  On the other hand, all three 
galaxies have  very different gas truncation radii.   NGC 4522 has a smaller gas truncation
radius \citep[0.35 $R_{25}$; ][]{kenney04} compared to that of NGC 4402 \citep[0.6 
$R_{25}$; ][]{crowl05} because a significant amount of \HI\ in NGC 4522 is 
extraplanar.  And since NGC 4501 is still in a pre-peak phase of ram pressure
stripping \citep{vollmer08}, its \HI\ disk is truncated within the stellar disk 
($\sim R_{25}$).

\begin{figure*}
\begin{center}
\includegraphics[scale=0.21]{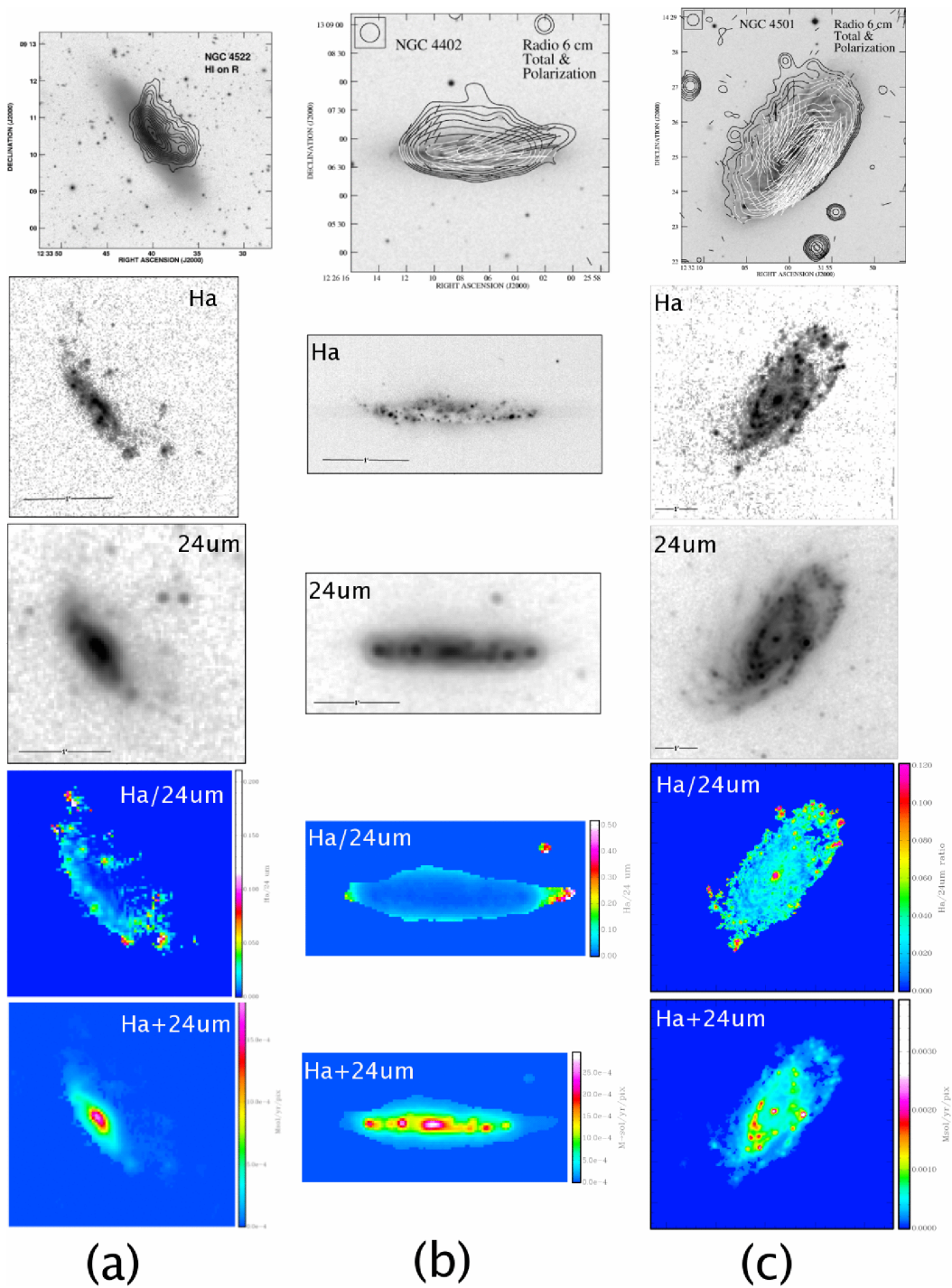}
\end{center}
\caption{\footnotesize{Columns (a), (b), (c) correspond to NGC 4522,
NGC 4402 and NGC 4501, respectively. Each columns consists of 5 panels;
which show  (from top to bottom) the grayscale $R$-band image overlaid
with \HI\ or radio continuum contours as well as polarization vectors, 
the H$\alpha$ image, the 24 $\mu$m image, the H$\alpha$$+24$ $\mu$m 
star formation map and the H$\alpha$$/24$ $\mu$m ratio map.}}
\label{oldimages}
\end{figure*}

The effect of the ICM--galaxy interaction is best illustrated by comparing 
the optical stellar morphology to the \HI\ or radio continuum morphology 
of each galaxy.  As shown in Figure~\ref{oldimages}, the \HI\  morphology of 
NGC 4522 is distorted and is shaped roughly like a bow-shock compared to the 
undisturbed stellar disk observed in the optical $R$-band \citep{kenney04}. 
The \HI\  morphology clearly shows the neutral gas being stripped out of the disk as the galaxy 
travels through the ICM.  Similarly, the radio continuum morphologies of 
NGC 4402 and NGC 4501 show compression on the sides of the leading edges 
as well as extended tails on the trailing sides \citep{crowl05,vollmer07,vollmer08}.  
Evidence of compression and shear are also shown by the total polarization 
vectors (marked by the white vectors in Figure~\ref{oldimages}).  

Figure~\ref{oldimages} presents the H$\alpha$ and 24 $\mu$m observations
for NGC 4522 (column a), NGC 4402 (column b) and NGC 4501 (column c). Each
column consists of five panels which  (from top to bottom) show (i) the grayscale 
$R$-band image overlaid with the \HI\ or radio continuum contours as well as 
polarization vectors, (ii) the H$\alpha$ map, (iii) the $24$ $\mu$m map,
(iv) the  H$\alpha$$/24$ $\mu$m ratio map, and (v) the total star formation 
rate map determined from the observed H$\alpha$$+24$ $\mu$m emission.

Section 2.1 describes the star-forming regions and the total star formation rate 
observed in the three galaxies and the ratios of the H$\alpha$ to 24 $\mu$m 
emission are discussed in Section 2.2.



\subsection{Star formation in NGC 4522, NGC 4402 and NGC 4501}
We aim to identify regions of ram pressure-induced star formation which
might plausibly be occuring since all three galaxies show evidence of active pressure.
Using the new 24 $\mu$m data from the Spitzer Survey of Virgo (SPITSOV; see 
Kenney et al.\ in these proceedings)  in combination with previous H$\alpha$ 
observations,  we examine the total star formation rate within these galaxies.  
The total star formation rate can be derived using a linear combination of the 
observed H$\alpha$ and 24 $\mu$m emission \citep{calzetti07}:
\begin{equation}
SFR (\mathrm{M}_{\odot} \; yr^{-1}) = 5.3 \times 10^{-42}\, [L(\mathrm{H}\alpha)+ 0.031\,L(24 \mu\mathrm{m})]
\end{equation}

\begin{table}
\footnotesize{
\caption{\footnotesize{Total  star formation rates of 
NGC 4402, NGC 4501 and NGC 4522.}}
\label{sfr}
\begin{center}
\begin{tabular}{lccc}
\tableline \tableline
Galaxy & Inclination &\multicolumn{2}{c}{SFR} \\
 & &H$\alpha$ & H$\alpha$+24 $\mu$m\\ 
\tableline
NGC 4402 & 80$^{\circ}$ & 0.28 M$_{\odot}$ & 0.60 M$_{\odot}$ \\
NGC 4501 & 62$^{\circ}$ & 1.67 M$_{\odot}$ & 2.54 M$_{\odot}$ \\
NGC 4522 & 79$^{\circ}$ & 0.10 M$_{\odot}$ & 0.16 M$_{\odot}$ \\
\tableline \tableline
\end{tabular}
\end{center}}
\end{table}

Estimates of the global star formation rates from the combination of 
H$\alpha$ and 24 $\mu$m emission were found to be 1.5, 1.6 and 2.1 times that of the 
star formation rates derived from the H$\alpha$ emission alone \citep[using the 
calibration from ][]{kennicutt98} for NGC 4501, NGC 4522 and NGC 4402,
respectively (see Table~\ref{sfr}).  The greatest difference is observed in 
the most inclined galaxy ($i=80^{\circ}$), NGC 4402.  Therefore, the difference
between the total star formation rates derived using the two methods
 appears to be correlated with the observed inclination of the galaxy.  This suggests
that more inclined galaxies with more dust will have a greater 24 $\mu$m 
contribution to the total star formation rate.

From the total star formation rate maps of all three galaxies (in Figure~\ref{oldimages}), 
the regions of most intense star formation are found in the galaxy centers and inner 
disk regions. We do not find any strongly-enhanced star-forming regions along the 
leading edges of interaction.  However, extraplanar star formation is clearly observed 
in NGC 4522 where stars have been formed from the stripped gas and are entering the
galaxy halo or intracluster space.  By comparing the H$\alpha$ and 24 $\mu$m fluxes 
in the  disk and extraplanar regions of NGC 4522, we find that $\sim$18\% of the total 
24 $\mu$m flux and $\sim$11\% of the total H$\alpha$ emission originated from the 
extraplanar regions of NGC 4522. This implies that the extraplanar star-forming regions 
are more obscured, on average, than the disk regions.  It is possible that the dust is 
being pushed between the observer and the star-forming regions as NGC 4522 travels away 
from the observer through the ICM.



\subsection{Spatial distribution of H$\alpha$ to 24 $\mu$m emission ratios}
For a galaxy moving towards the observer, one might expect ram pressure
to sweep the dust behind the observed star-forming regions causing a lower 
H$\alpha$ attenuation. Conversely, a galaxy moving 
away from the observer may display a higher H$\alpha$ attenuation as more 
stripped dust is pushed between the observer and the star-forming regions.
Using the H$\alpha$/24 $\mu$m ratios, we investigate the spatial distribution
of H$\alpha$ attenuation in our sample of cluster galaxies where ram pressure
is active. It should be noted that both NGC 4501 and NGC 4522 are moving
away from the observer, while NGC 4402 is moving towards the observer.

As expected, the most obscured regions of star formation are found in the 
galaxy centers (see the H$\alpha$/24 $\mu$m ratio maps in Figure~\ref{oldimages}), 
whereas the least obscured star-forming regions are found in the outer edges of all 
three galaxies. In particular, regions of high H$\alpha$/24 $\mu$m ratios are observed 
(from Figure~\ref{oldimages}) near the leading edge of interaction of NGC 4522.   

The global H$\alpha$$/24$ $\mu$m ratios of our three galaxies  are smaller
 than the median H$\alpha$$/24$ $\mu$m ratio (0.034) of the SINGS sample used 
by \citet{calzetti07}.  We find H$\alpha$$/24$ $\mu$m ratios of 0.029, 0.023 and 
0.008 for NGC 4501, NGC 4522 and NGC 4402, respectively.  It should be noted that
the SINGS sample used by \citet{calzetti07} consists of galaxies spanning the full 
range of inclinations, while our galaxies have  inclinations ranging from $62^{\circ}$ 
to $80^{\circ}$. Our current results suggest that a  greater inclination may
 be correlated with a smaller global H$\alpha$$/24$ $\mu$m ratio.

\begin{figure*}
\begin{center}
\includegraphics[scale=0.175]{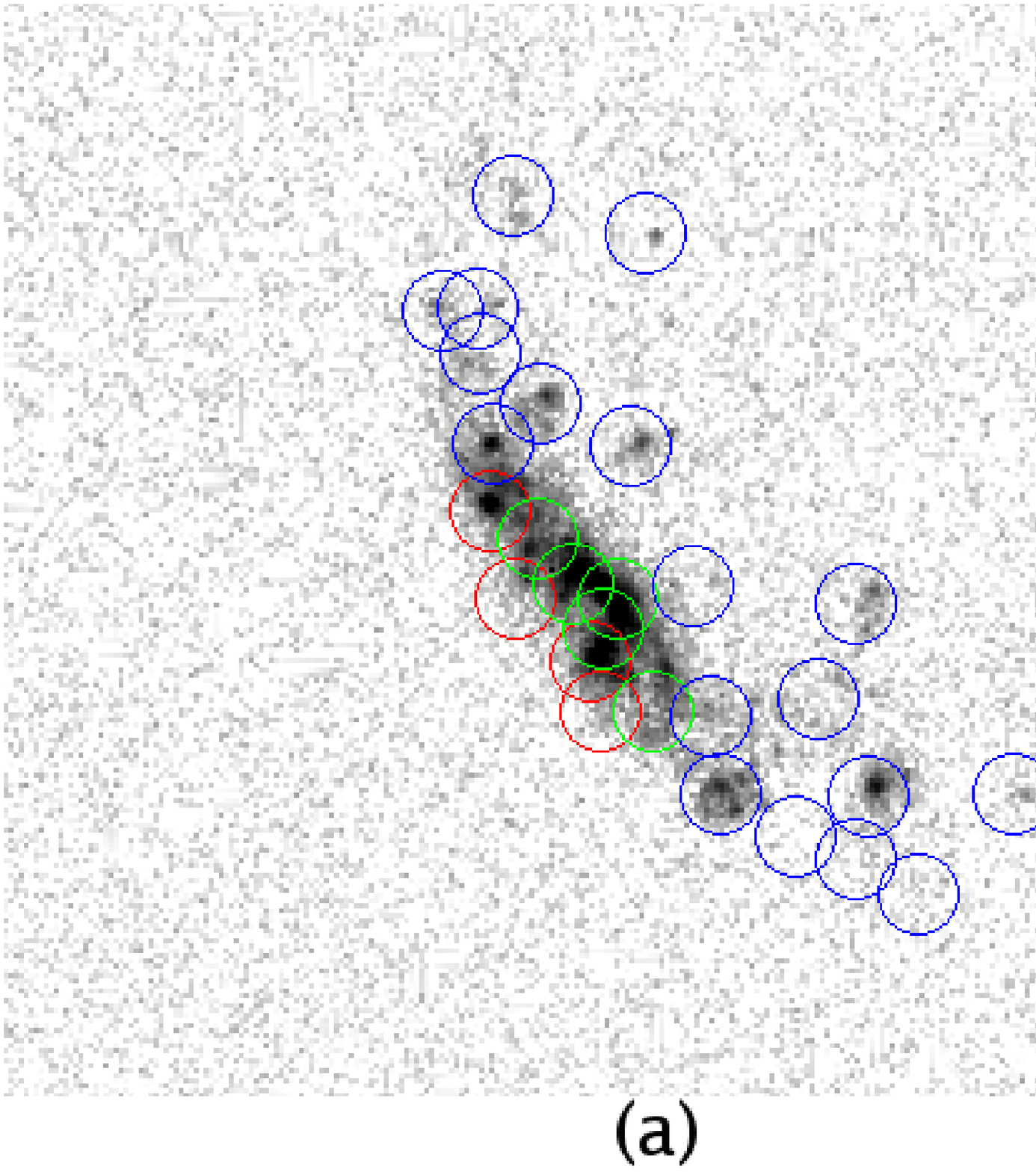}
\end{center}
\caption{\footnotesize{ The 24 $\mu$m/H$\alpha$ ratio of  500 pc regions
in NGC 4522.  The H$\alpha$  and 24 $\mu$m observations overlaid with 
regions of 500 pc diameter apertures are shown in panels (a) and (b), 
respectively. The regions are divided into three categories; regions within 
the inner disk (green), regions on the leading edge of the ISM--ICM interaction 
(red) and outer disk or extraplanar regions (blue).  Panel (c) shows the  
24 $\mu$m/H$\alpha$ ratio of each 500 pc region as a function of the 
projected radius from the center of the galaxy.  The regions circled 
in blue, red and green correspond to the blue, red and green dots in the
right panel respectively.}}
\label{n4522}
\end{figure*}
\begin{figure*}
\begin{center}
\includegraphics[scale=0.575]{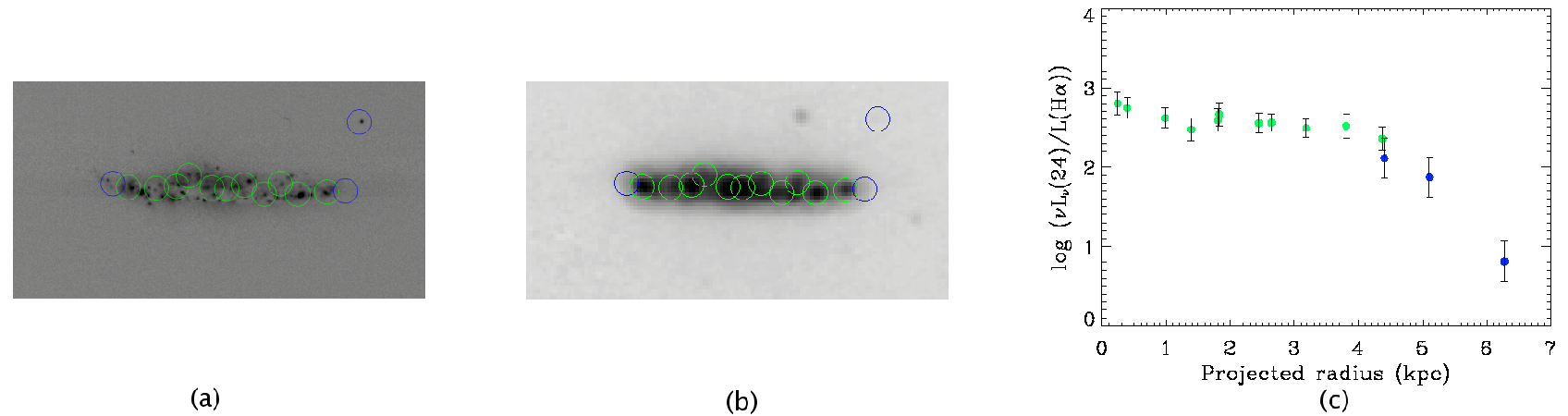}
\end{center}
\caption{\footnotesize{ The 24 $\mu$m/H$\alpha$ ratio of  500 pc regions
in NGC 4402. See figure~\ref{n4522} for more details.}}
\label{n4402}
\end{figure*}

\begin{figure*}
\begin{center}
\includegraphics[scale=0.208]{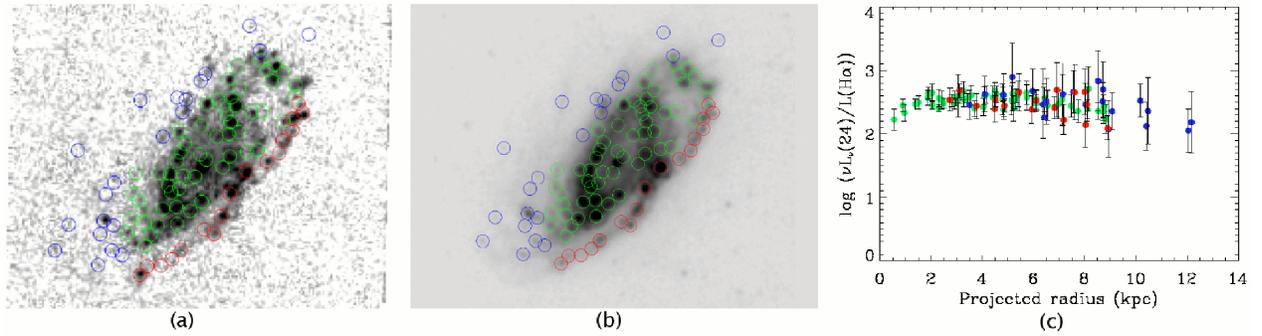}
\end{center}
\caption{\footnotesize{The 24 $\mu$m/H$\alpha$ ratio of  500 pc regions
in NGC 4501. See figure~\ref{n4522} for more details.}}
\label{n4501}
\end{figure*}

Recently, \citet{prescott07} found decreasing radial trends in H$\alpha$ 
attenuation by dust for a majority of the SINGS galaxies. Similar to \citet{prescott07}, 
we selected regions with circular apertures of 500 pc and measured the ratios of 
$\nu L_{\nu}(24)/L(H\alpha)$ as a function of the projected radius from the centers 
of our galaxies.  However unlike \citet{prescott07}, our regions were selected 
from the H$\alpha$ observations as well as the 24 $\mu$m observations in order 
to probe the outer star-forming regions more effectively and avoid biasing against 
very unobscured regions.  Although an aperture size of 500 pc is not optimal for 
separating individual star-forming regions, it is comparable to the spatial 
resolution of the 24 $\mu$m observations. 

Figures~\ref{n4522},~\ref{n4402} and~\ref{n4501} show the 
regions selected in  NGC 4522, NGC 4402 and NGC 4501, 
respectively.  The first two panels of each figure (from the left) show 
the H$\alpha$  and the 24 $\mu$m map of the galaxy with the selected 
regions overlaid.  The final panel of each figure (c) plots  
$\nu L_{\nu}(24)/L(H\alpha)$ as a function of projected radius from the 
galaxy center. The selected regions are loosely divided 
into three categories; regions on the side of the leading edge (red),
extraplanar or outer disk regions (blue) and regions in the inner
disk (green).   These categories are color-coded for easier
identification and the color of the apertures shown on the maps correspond 
directly to the color of the dots on the plot in panel (c).

The average 24 $\mu$m/H$\alpha$ ratios measured from the inner disk regions (green 
regions) of the three galaxies are greater than those measured from the 
outer disk or extraplanar regions (blue regions). The average ratios for the 
inner disk regions of all three galaxies are very similar (with values between 2.52
and 2.64) but the difference in the 24 $\mu$m/H$\alpha$ ratio between the inner
and outer regions varies significantly for each galaxy.  The minimum difference
in average ratios between the inner and outer disk is $\sim1.6\%$ for NGC 4501 and
the maximum difference between the outer and inner disk is $\sim22.9\%$ for NGC 4402.
Since NGC 4402 is highly inclined, the 
24 $\mu$m observations will be biased by the dusty disk of the galaxy.  However,
the motion of NGC 4402 towards the observer is able to push the dust from 
the outer disk (or extraplanar) regions behind star-forming regions as these
regions are not as affected by the dusty center.  This may explain the fact that 
a decrease in the 24 $\mu$m/H$\alpha$ ratio is observed only at the two extreme ends
and the extraplanar \HII\ region of  NGC 4402.

Of the three galaxies analysed, NGC 4402 exhibits the most convincing radial 
decrease in the 24 $\mu$m/H$\alpha$ ratio (see Figure~\ref{n4402}). Its 
24 $\mu$m/H$\alpha$ ratio is highest in the central 0.5 kpc and then decreases 
very gradually to 4 kpc. The two outer disk regions on both sides of the disk 
(as represented by the two blue points at radius $\sim$4--5 kpc in panel (c) of 
Figure~\ref{n4402}) exhibit a clear decrease in the 24 $\mu$m/H$\alpha$ ratios.
In addition, the region at the largest projected radius is an \HII\ region, 
{\em{122603+130724}},  which appears to be the {\em{most unobscured}} star-forming 
region found in the current analysis. This extraplanar \HII\ region is thought to 
have formed  from enriched material stripped from NGC 4402 \citep{cortese04}.  
It should be noted that these three outermost regions in NGC 4402 have fairly low 
H$\alpha$ luminosities (i.e.\ they are not high luminosity \HII\ regions with low 
obscurity).

Even though bright star-forming regions have been observed near the leading 
edge in the optical images of NGC 4402 \citep{crowl05}, the 24 $\mu$m/H$\alpha$ 
ratios in this part of the galaxy do not appear unusual. The 
24 $\mu$m/H$\alpha$ ratios of NGC 4402 are comparable to the the ratios 
obtained by \citet{prescott07} for the highly-inclined SINGS sample.  The lowest
log $[\nu L_{\nu}(24)/L(H\alpha)]$ ratio found by \citet{prescott07} for the 
highly-inclined SINGS sample is $\sim$0.6, while the most unobscured 
extraplanar \HII\ region in NGC 4402 is $\sim$0.8.  On scales of 500 pc, it is 
not possible to identify leading edge regions in this galaxy due to resolution constraints.

Unlike NGC 4402, we do not find significant radial decreases in the H$\alpha$ 
attenuation for NGC 4501 and NGC 4522.  A very slight radial decrease in the 
H$\alpha$ attenuation  is observed in NGC 4522 (Figure~\ref{n4522}) but there 
remains a large scatter in the ratios measured from the outer disk (or extraplanar) 
regions. Unlike Figure~\ref{oldimages}, we do not find any evidence for 
highly-unobscured star formation along the leading edge of NGC 4522 in 
Figure~\ref{n4522}.  This may be due to the fact that the regions of low H$\alpha$ 
attenuation in NGC 4522  are located in small regions with low luminosities.
Consequently, the effect of decreased H$\alpha$ attenuation from these regions 
is diminished when the ratios are averaged across 500 pc apertures. In addition,
the obscurity of the extraplanar regions of this galaxy is enhanced due to its
direction of motion away from the observer through the ICM. 

Similar to NGC 4522, only a slight decrease is found in NGC 4501 (Figure~\ref{n4501}).   
There are two possible explanations which may account for the lack of an observed
decrease in the H$\alpha$ attenuation at the leading edge.  Firstly, NGC 4501 is 
travelling through the ICM away from the observer which will result in a displacement 
of the gas and dust from the regions close to the leading edge towards the observer. 
Secondly,   \citet{vollmer08} concluded that NGC 4501 is in a pre-peak 
stripping stage  and that ram pressure will only reach its maximum in $\sim100$ Myr.
Hence,   it is possible that the current ram pressure is not strong enough
to remove enough dust at the leading edge even though there is enough
pressure to strip off the diffuse gas.

\section{Summary}
Using the 24 $\mu$m observations obtained from the Spitzer Survey of Virgo (SPITSOV)
in combination with H$\alpha$ observations, we explored the 
distribution of star-forming regions within three Virgo Cluster galaxies (NGC 
4402, NGC 4501 and NGC 4522) which are experiencing active ram pressure.
The combination of the 24 $\mu$m and H$\alpha$ observations allowed us
 to uncover regions of significant dust obscuration.  From this we are able to 
produce total star formation maps without the usual bias caused by dust extinction.  

From our preliminary analysis of NGC 4522, NGC 4402 and NGC 4501, we 
find evidence neither for highly obscured star forming regions nor for 
strongly enhanced star formation along the leading edges of interaction.  The
H$\alpha$ attenuation (ratio between H$\alpha$ and 24 $\mu$m emission) of these 
three galaxies are comparable to the non-cluster SINGS galaxies studied by 
\citet{calzetti07} and \citet{prescott07}.  However, NGC 4402 exhibits
unobscured outer disk and extraplanar star-forming regions, while, the outer 
disk and extraplanar star-forming regions of NGC 4522 appears moderately
obscured.  We attribute the differences between these two galaxies to the direction
of motion of each galaxy through the ICM.

\acknowledgments
This work is based on observations made with the Spitzer Space Telescope, which is 
operated by the Jet Propulsion Laboratory, California Institute of Technology 
under a contract with NASA. Support for this work was provided by NASA through 
an award issued by JPL/Caltech.




\end{document}